\documentclass[10pt,letterpaper]{article}
\usepackage{opex3}
\usepackage{color}
\usepackage{amsmath}
\usepackage{graphicx}
\usepackage{dsfont}
\usepackage[mathscr]{euscript}
\usepackage{amssymb}
\usepackage{mathtools}
\usepackage{cite}
\newcommand{\defeq}{\vcentcolon=}

\begin{document}

\title{Fast Hadamard transforms for compressive sensing of joint systems: measurement 
of a 3.2 million-dimensional bi-photon probability distribution}

\author{Daniel J. Lum,$^{1,2,*}$ Samuel H. Knarr,$^{1,2}$ and John C. Howell$^{1,2}$}

\address{$^{1}$Department of Physics and Astronomy, University of Rochester, Rochester, New York 14627, USA\\
$^{2}$Center for Coherence and Quantum Optics, University of Rochester, Rochester, New York 14627, USA\\}
\email{$^*$daniel.lum@pas.rochester.edu}

\begin{abstract}
We demonstrate how to efficiently implement extremely high-dimensional compressive imaging of a bi-photon probability distribution. Our method uses fast-Hadamard-transform Kronecker-based compressive sensing to acquire the joint space distribution. We list, in detail, the operations necessary to enable fast-transform-based matrix-vector operations in the joint space to reconstruct a 16.8 million-dimensional image in less than 10 minutes. Within a subspace of that image exists a 3.2 million-dimensional bi-photon probability distribution. In addition, we demonstrate how the marginal distributions can aid in the accuracy of joint space distribution reconstructions.
\end{abstract}


\section{Introduction}
Characterizing high-dimensional joint systems is a difficult problem due to experimental impracticalities such as long measurement times, low flux, or insufficient computing resources. One example of such a characterization is of continuous-variable entangled states -- a resource gaining ground in quantum technologies \cite{masada2015continuous,o2009photonic,braunstein2005quantum,steane1998quantum,huver2008entangled,giovannetti2011advances}. A widely used source of continuous-variable entangled states is Spontaneous Parametric Down-Conversion (SPDC) in a nonlinear crystal \cite{rubin1996}. Depending on the configuration, the resulting bi-photon state may be entangled in the transverse degrees of freedom \cite{howell2004,keller1997,schneeloch2015introduction}. To determine if the system is entangled, both the bi-photon joint position and joint momentum probability distributions composed of signal and idler photons must be measured through correlation measurements.
 
Much work has been done recently to characterize high-dimensional position-momentum entanglement with discrete measurements \cite{howell2004,schneeloch2013violation,schneeloch2014demonstrating,lvovsky2009continuous,giovannini2013,bolduc2015direct,shabani2011efficient}. Characterizations of position or momentum distributions resulting from SPDC are done by measuring signal and idler pixel correlations in either an image plane of the crystal (constituting a position measurement) or a Fourier-transform plane of the crystal (constituting a momentum measurement) through coincidence counting.
 
For a bi-photon probability distribution, measurements are typically done by measuring correlations via raster scanning through the individual signal- and idler-photon probability distributions. The time required to complete a raster scan with single-photon detectors quickly becomes impractical for certain scans.  Imaging these distributions with a camera has been shown in \cite{edgar2012imaging,fickler2013real}, yet cameras often introduce far more noise than a single-photon detector.
 
Recently, Compressive Sensing (CS) \cite{donoho2006compressed,duarte2008single} techniques have been introduced as an alternative to raster scanning for characterizing a high-dimensional entangled system, dropping the measurement time from months to hours \cite{howland2013efficient,tonolini2014}. While the data-acquisition time is drastically reduced, it comes at the cost of computational complexity, requiring a computational reconstruction of the signal. Performing CS on high-dimensional signals is not a new problem, and several clever solutions exist for utilizing separable compressive sensing matrices combined by a Kronecker product \cite{rivenson2009practical,duarte2012kronecker}. However, these methods are ill suited for sampling the correlations in a joint space.
 
In this article we propose the use of fast Hadamard transforms for high-dimensional joint space reconstructions. Specifically, we show how the Kronecker-product-based recursion relations of Sylvester-type Hadamard matrices can combine single-particle sensing matrices. This, in turn, enables the use of fast Hadamard transforms in the joint space as they have been shown to drastically reduce CS reconstruction times \cite{shishkin2012fast}. Using the randomization techniques outlined in \cite{li2011compressive,li2009user}, sensing matrices composed of randomized Hadamard matrices offer tremendous speed enhancements in many reconstruction algorithms.
 
Additionally we show that by using the individual signal and idler marginal distributions in our reconstruction of the joint space distribution, we can more accurately acquire transverse spatial correlations as measured by the mutual information. To the best of our knowledge, this is the first time the single particle information has been used in reconstructing the joint space distribution.
 
To demonstrate the effectiveness of our method, reconstruct a $16.8\times10^6$ dimensional joint-space distribution in only a few minutes. Within this joint-space lives a 3.2 million-dimensional bi-photon position probability distribution from which we measure the degree of transverse correlations between signal and idler photons using the mutual information. 

The experimental realization here is closely related to the work performed in \cite{howland2013efficient}. The experiment in this article is merely meant to demonstrate how structured randomness enables efficient reconstructions of the joint space distribution at even higher dimensions. In theory, this increase in resolution allows for an increase in the amount of measurable mutual information.

\section{Compressive sensing} CS helps to overcome unreasonable data-acquisition times 
associated with sampling signals with limited resources. CS requires 
that the signal of interest $\mathbf{\stackrel{\sim}{x}}$ has a sparse representation 
$\mathbf{x}$ via a basis transform. Limiting the discussion to real signals for 
simplicity, $\mathbf{\stackrel{\sim}{x}}\in\mathbb{R}^N$, then there 
exists a basis transform $\Psi$ in which $k < N$ components of $\mathbf{x} = 
\Psi\cdot\mathbf{\stackrel{\sim}{x}}$ are nonzero; $\mathbf{x}$ is a $k$-sparse 
representation of $\mathbf{\stackrel{\sim}{x}}$. CS posits that if $\mathbf{x}$ is 
approximately $k$-sparse, then only $M = \mathscr{O}\left(k\log(N/k)\right)$ projections of 
$\mathbf{\stackrel{\sim}{x}}$ are needed to accurately sample 
$\mathbf{\stackrel{\sim}{x}}$. A typical CS technique involves taking $M \ll N$ projections 
of a signal $\mathbf{\stackrel{\sim}{x}}$ with a random sensing matrix 
$\mathbf{A}\in\mathbb{R}^{M\times N}$ to form a measurement vector $\mathbf{y} = 
\mathbf{A}\cdot\mathbf{\stackrel{\sim}{x}}$ where $\mathbf{y}\in\mathbb{R}^M$. Assuming 
the signal representation $\mathbf{x}$ is $k$-sparse, $\mathbf{x}$ may be recovered by 
solving 
\begin{equation}\label{eq:cs} \min_{\mathbf{x}} \, \tau g(\mathbf{x}) \qquad \text{ 
subject to }\qquad \|\mathbf{y}-\mathbf{A}\cdot \left(\Psi^{-1}\cdot \mathbf{x}\right)\|_2^2 < 
\epsilon ,
\end{equation} where the second term is a least-squares term bounded by a 
predefined error $\epsilon$ and 
\begin{equation*}
 \|\mathbf{x}\|_p \defeq \left(\sum\limits_{i=1}^n |x_i|^p\right)^{1/p}
\end{equation*} 
is defined as the $l_p$-norm. The function $g(\mathbf{x})$ is a function 
to be minimized depending on the assumed sparsity; $g(\mathbf{x})$ is often chosen to be 
the $l_1$-norm $\|\mathbf{x}\|_1^2$, the total variation as defined by the gradient 
$\|\nabla\mathbf{x}\|_1$, or a combination of functions where $\mathbf{x}$ is known to be 
sparse. $\tau$ is simply a weighting term that weights the sparsity, favoring either the 
least-squares solution or the sparsity. Because $\mathbf{A}\in\mathbb{R}^{M\times N}$ 
where $M\ll N$, there are an infinite number of solutions confined to the least-squares 
term. The function $g(\mathbf{x})$ picks out the sparsest of these solutions which should correspond to our signal $\mathbf{x}$. An overview 
of compressive sensing and its applications may be found in 
\cite{candes2008introduction}.

\section{Measuring a non-separable joint system}

We apply CS to measure the joint position probability distribution of the down-converted signal and idler photons from SPDC. Quantum mechanics tells us that the bi-photon state exists in a Hilbert space composed of the tensor product of the individual signal and idler photon Hilbert spaces. After representing the bi-photon state in a discrete basis, operators on these states are represented as matrices. In order make a measurement, we approximate the state as living in a finite dimensional Hilbert space. We can therefore represent a bi-photon operator matrix in terms of Kronecker products of individual signal and idler photon operator matrices. For CS, we can let these operators be projection operators and manipulate them such that they form the rows of a sensing matrix $\mathbf{A}$. The sets of projection operators for the signal and idler spaces are designated by a a set of $M$ patterns where each pattern is in $\mathbb{R}^N$; $\mathbf{P}_S$ and $\mathbf{P}_I \in\mathbb{R}^{M\times N}$. In this manner, the sensing matrix is written as
\begin{equation} \label{eq:sensing_matrix} 
\mathbf{A} =  \left[ \begin{array}{c} \mathbf{P}_S[1] \otimes \mathbf{P}_I[1] \\ 
\mathbf{P}_S[2] \otimes \mathbf{P}_I[2] \\ \vdots \\ \mathbf{P}_S[M] \otimes 
\mathbf{P}_I[M] \end{array} \right] \end{equation} for $i \in 1...M$ where 
$\mathbf{P}[i]$ represents the $i^{th}$ row of $\mathbf{P}$.
 
The Kronecker product $\otimes$ in this article operates on matrices and vectors such that if $\mathbf{a}$ is of dimension $m\times n$ and $\mathbf{b}$ is of dimension $p\times q$, then their Kronecker is of dimension $mp\times nq$ represented as
\begin{equation}
\mathbf{a}\otimes\mathbf{b} = 
\left[ 
\begin{array}{ccc} 
a_{11}\mathbf{b} & \cdots & a_{1n}\mathbf{b} \\ 
\vdots & \ddots & \vdots \\ 
a_{m1}\mathbf{b} & \cdots & a_{mn}\mathbf{b}
\end{array}
\right].
\end{equation}

\begin{figure} \includegraphics[width=1\textwidth]{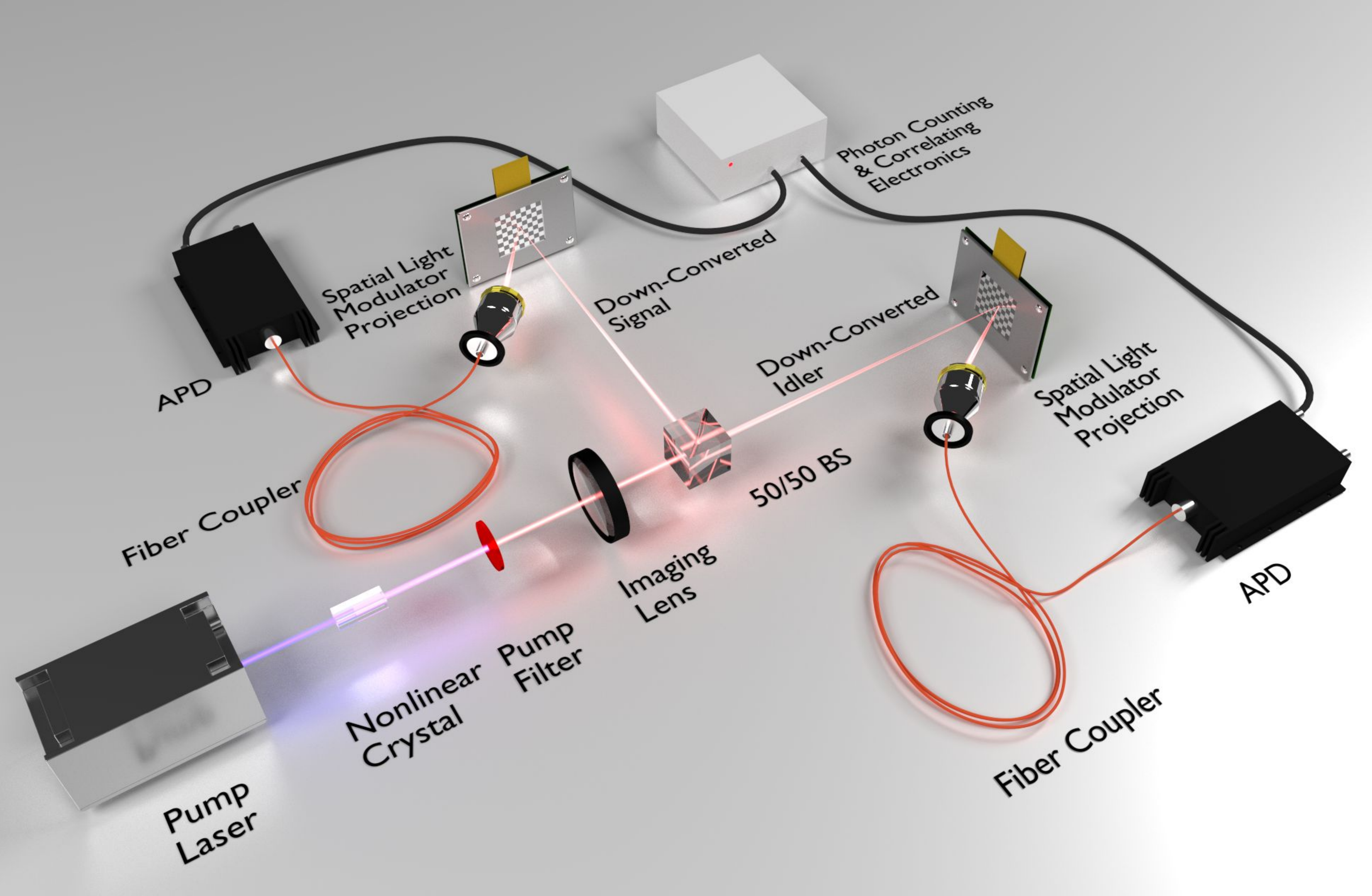} \caption{The above 
experimental diagram demonstrates how to image a joint two-particle system. In this 
paper, the joint system is composed of highly-correlated signal and idler photons from a 
SPDC source. The experiment samples the position distribution of the joint system by 
taking random projections of signal and idler intensities with a spatial light modulator 
within an image plane of the crystal. An avalanche photodiode (APD) detects photon 
arrivals while the photon counters measure photon coincidences.} \label{fig:Experiment} 
\end{figure}

An experimental 
diagram for measuring the joint space bi-photon position probability distribution is presented in 
Fig. \ref{fig:Experiment} where each particle's space is depicted as two-dimensional. Yet, to simplify 
the CS formalism, we represent the signal and idler spaces as one-dimensional living in $\mathbb{R}^N$ and the 
joint space distribution as a vector $\mathbf{x}\in\mathbb{R}^{N^2}$. As outlined in \cite{howland2013efficient}, compressive sensing is experimentally 
accomplished by taking random projections with patterns composed of 
$\left[\sqrt{N}\times\sqrt{N}\right]$ pixels within each subspace of the signal and idler 
systems and then measuring the resulting correlations in photon counts.

During reconstruction, the bi-photon probability 
distribution is already sparse in the pixel basis due to the tight pixel 
correlations resulting from energy and momentum conservation. This eliminates the need to define a sparse-basis such that $\Psi$ becomes 
the identity operator $\Psi = \mathds{1}$ in the formalism above.

Random binary matrices were used in \cite{howland2013efficient}, yet we wish to use \emph{structured} random binary matrices for reconstruction purposes. Using properties of Kronecker products enables relatively efficient computations of the reconstruction operations $\mathbf{A}\cdot\mathbf{x}$ and $\mathbf{A}^{T}\cdot\mathbf{y}$, where $\mathbf{A}^{T}$ is the transpose of $\mathbf{A}$, because $\mathbf{A}$ never needs to be computed explicitly \cite{horn1991matrix}. However, structured randomness can enable the use of fast transforms which are even more efficient. Our contribution is to demonstrate in the following section how randomized Sylvester-Hadamard matrices enable the use of fast Hadamard transforms in joint space CS reconstructions.

\section{Fast Hadamard transform based sensing matrices}

\subsection{Randomly sampled \& permuted Hadamard sensing matrices}

Sylvester-Hadamard matrices have a structure that is particularly advantageous to the CS framework. 
These matrices are generated from a simple recursion relation defined by a Kronecker product.
\begin{eqnarray*} \mathbf{H}_1 & = & [1] \\ \mathbf{H}_2 & = & \left[ \begin{array}{cc} 1 
& 1 \\ 1 & -1 \end{array} \right].\\ \end{eqnarray*} From these, any Sylvester-Hadamard
matrix can be decomposed as follows:
\begin{equation} 
\label{eq:kronecker}
\mathbf{H}_{2^k} = 
\mathbf{H}_2 \otimes 
\mathbf{H}_{2^{k-1}} =
\left[ 
\begin{array}{cc} 
\mathbf{H}_{2^{k-1}} & \mathbf{H}_{2^{k-1}} \\ 
\mathbf{H}_{2^{k-1}} & - \mathbf{H}_{2^{k-1}} 
\end{array} \right].
\end{equation}
for $k > 1$. Because of this structure, Sylvester-Hadamard matrices are restricted to powers of two but can be used to build patterns and a sensing matrix that utilizes the speed and 
efficiency of a fast Hadamard transform $\mathscr{H}[...]$. We use a normally-ordered 
fast transform in this work. Its algorithm is similar to that of a fast Fourier 
transform, but it consists of only additions and subtractions. Hence, it performs the reconstruction operations $\mathbf{A}\cdot 
\mathbf{x}$ and $\mathbf{A}^T\cdot \mathbf{y}$ in $\mathscr{O}\left(N^2\log N 
\right)$ time -- significantly faster than an explicit matrix-vector multiplication of $\mathscr{O}\left(N^4\right)$ 
time. A thorough overview of Hadamard matrices, fast-Hadamard 
transforms, and their applications to signal and image processing may be found in 
\cite{yarlagadda1997hadamard}.

To construct $\mathbf{P}_S$, $\mathbf{P}_I$, and $\mathbf{A}$ from Hadamard matrices, the 
Hadamard matrices must be randomized in both their rows and columns. 
The sensing matrices 
must be both incoherent with the image yet span the space in which the signal resides. In 
other words, the sensing matrices must adequately sample the basis components of the 
signal. Random sensing matrices perform this task well, yet Hadamard matrices 
naturally contain much structure.
To begin, each sensing matrix must be formed by taking specific rows from a 
Hadamard matrix with the correct dimensions. $\mathbf{A}$ is constructed from 
$\mathbf{H}_{N^2}$ while $\mathbf{P}_S$ and $\mathbf{P}_I$ are constructed from 
$\mathbf{H}_{N}$. Because of the relation in Eq. (\ref{eq:sensing_matrix}), the rows 
of $\mathbf{A}$ will be determined by the rows of $\mathbf{P}_S$ and $\mathbf{P}_I$.

The randomization of the Hadamard matrix rows is accomplished by
defining two vectors $\mathbf{r}_S$ and $\mathbf{r}_I \in\mathbb{R}^{M}$ for each signal 
and idler system composed of $M$ randomly chosen integers on the interval [2,N]. 
The values in $\mathbf{r}$ state which rows should be extracted from $\mathbf{H}_N$ when constructing $\mathbf{P}_S$ and $\mathbf{P}_I$. Note 
that the interval begins at 2 because the first row of a Hadamard matrix is composed 
entirely of ones. The interval may begin at 1 if the total photon flux on a detector is desired.
Also, note that 
$\mathbf{A}\in\mathbb{R}^{M\times N^2}$ where $M << N^2$. This condition allows for 
scenarios where $\mathbf{P}_S, \mathbf{P}_I\in\mathbb{R}^{M\times N}$ such that $M > N$ 
in the individual subspaces, meaning rows of $\mathbf{H}_N$ may be repeated when 
constructing $\mathbf{P}_S$ and $\mathbf{P}_I$.

The randomization of the Hadamard columns is accomplished by defining permutation vectors 
$\mathbf{p}_S$ and $\mathbf{p}_I \in\mathbb{R}^N$ that randomly permute the $N$ columns 
of $\mathbf{H}_N$. Once $\mathbf{r}$ and $\mathbf{p}$ have been defined for both the 
signal and idler subspaces, patterns are constructed by the following equations: 
\begin{eqnarray} \label{eq:Patterns} \mathbf{P}_S & = & 
\mathbf{H}_{N}[\mathbf{r}_S,\mathbf{p}_S ] \nonumber \\ \mathbf{P}_I & = & 
\mathbf{H}_{N}[\mathbf{r}_I,\mathbf{p}_I] \end{eqnarray} where the $y$ and $x$ components 
of $\mathbf{H}[y,x]$ refer to the rows and columns of $\mathbf{H}$ respectively.

Although these operations are defined for 
the signal and idler subspaces, they combine in a particular way according to Eq. 
(\ref{eq:sensing_matrix}) to construct a Hadamard-based sensing matrix $\mathbf{A}$ that 
enables fast transform operations in the joint space. The manner in which they combine to 
manipulate a Hadamard matrix $\mathbf{H}_{N^2}$ that spans the joint space is detailed in 
the next section.

\subsection{Joint space Sylvester-Hadamard sensing matrices} 
Once $\mathbf{r}$ and $\mathbf{p}$ have been defined for the individual signal and idler subspaces, they may be used to construct 
the corresponding joint space row-selection and permutation 
vectors, $\mathbf{r}_{SI}$ and $\mathbf{p}_{SI}$. Consider the construction of $\mathbf{r}_{SI}$ first.  By Eq. (\ref{eq:kronecker}), 
the complete joint space sensing matrix $\mathbf{A}$ is simply formed by the row-wise 
Kronecker product of the subspace sensing matrices $\mathbf{P}_S$ and $\mathbf{P}_I$. As 
$\mathbf{r}_S$ and $\mathbf{r}_I$ determine the ordering of the Hadamard rows within 
these patterns, $\mathbf{r}_{SI}$ must also be a subset of a Kronecker product of 
$\mathbf{r}_S$ and $\mathbf{r}_I$.  Knowing that the Kronecker product of $\mathbf{P}_S$ 
and $\mathbf{P}_I$ will form ``blocks" of size $[M\times N]$, it is straightforward to 
show that 
\begin{equation}\label{eq:PicksN^2} \mathbf{r}_{SI}[i] = 
N\left(\mathbf{r}_S[i]-1\right)+\mathbf{r}_I[i] 
\end{equation} 
for $i \in 1...M$ where 
$\mathbf{r}[i]$ represents the $i^{th}$ component of $\mathbf{r}$. Note that element-wise 
counting in this article starts at 1 and not 0.

Because $\mathbf{r}_S$ and $\mathbf{r}_I$ are chosen at random and will often be 
over-complete, $M > N$, and $\mathbf{r}_{SI}$ will probably have repeating units, and a row 
within $\mathbf{A}$ will appear more than once. This is equivalent to taking the same 
projection more than once, offering no additional information. To prevent this, we simply compare each of the values within $\mathbf{r}_{SI}$ and eliminate 
repeating $i^{th}$ values. If $\mathbf{r}_{SI}[i]$ is a repeated value, we eliminate 
$\mathbf{r}_{SI}[i]$ along with the components $\mathbf{r}_S[i]$ and $\mathbf{r}_I[i]$. 
In this way, the number of samples $M$ will decrease yet contain the same amount of 
information.

The formation of $\mathbf{p}_{SI}$ follows a similar form as $\mathbf{r}_{SI}$, yet it 
will be of length $N^2$. Although it is not a simple Kronecker product, it does follow 
from the structure in Eq. (\ref{eq:sensing_matrix}). The structure of 
$\mathbf{p}_{SI}$ takes the form 
\begin{equation}\label{eq:PermN^2} 
\mathbf{p}_{SI}[N(i-1)+j] = N\left(\mathbf{p}_S[i]-1\right)+\mathbf{p}_I[j] 
\end{equation}
for $i \in 1...N$ and $j \in 1...N$. Generating randomized Hadamard matrices using 
$\mathbf{r}$ and $\mathbf{p}$ for each signal, idler, and joint space are summarized 
below: 
\begin{eqnarray} \label{eq:PatternsA} \mathbf{P}_S & = & 
\mathbf{H}_{N}[\mathbf{r}_S,\mathbf{p}_S ] \nonumber\\ \label{eq:Patterns} \mathbf{P}_I 
& = & \mathbf{H}_{N}[\mathbf{r}_I,\mathbf{p}_I]\\ \label{eq:A} \mathbf{A} & = & 
\mathbf{H}_{N^2}[\mathbf{r}_{SI},\mathbf{p}_{SI}] \nonumber \end{eqnarray} 
where 
the $y$ and $x$ components of $\mathbf{H}[y,x]$ refer to the rows and columns of 
$\mathbf{H}$ respectively. The construction of $\mathbf{A}$ presented in Eq. (\ref{eq:PatternsA}) allows us to use fast transforms as explained in the next section.

\subsection{Joint space fast Hadamard transform operations}

Keeping track of the randomization operations allows the use of fast Hadamard transforms when computing $\mathbf{A}\cdot\mathbf{x}$ and $\mathbf{A}^T\cdot\mathbf{y}$.
This is accomplished by reordering either $\mathbf{x}$ or $\mathbf{y}$ according to $\mathbf{p}$, taking the fast Hadamard 
transform, and then picking specific elements from the final 
result according to $\mathbf{r}$. The manner in which they are rearranged and picked 
depends upon the operation $\mathbf{A}\cdot \mathbf{x}$ or $\mathbf{A}^T\cdot \mathbf{y}$ 
in either the data acquisition or reconstruction processes.

Starting with the data-taking procedure $\mathbf{y} = \mathbf{A}\cdot \mathbf{x}$, 
projections are taken in each signal and idler system by first constructing individual patterns. Pattern construction is done by fast Hadamard transforming basis vectors and then 
permuting them. Because of the symmetric nature of a Hadamard matrix ($\mathbf{H} = \mathbf{H}^T$), a fast Hadamard 
transform of a basis vector $\boldsymbol{\alpha}[i]$, in which the $i^{th}$ component is 
equal to one and the rest zeros, is equal to the $i^{th}$ row of the Hadamard matrix 
$\mathbf{H}[i,:]$. In short, $\mathscr{H}\left[\boldsymbol{\alpha}[i]\right] = 
\mathbf{H}[i]$. Hence, every $i^{th}$ pattern $\mathbf{P}[i,:]$ can be built according 
to a fast transform by 
\begin{eqnarray} \label{eq:fastPattern}
\mathbf{H}[\mathbf{r}\left[i\right],:] &=& 
\mathscr{H}\left[\boldsymbol{\alpha}\left[\mathbf{r}[i]\right]\right] \nonumber \\ 
\mathbf{P}[i] &=& \mathbf{H}[\mathbf{r}\left[i\right],\mathbf{p}] \end{eqnarray} for $i \in 1...M$ where 
$\boldsymbol{\alpha}[\mathbf{r}[i]]$, a basis vector whose $\mathbf{r}[i]^{th}$ component 
is equal to 1, is fast-Hadamard transformed and then permuted according to $\mathbf{p}$.

To take experimental data, many SLM's, such as digital micromirror devices, are operated 
in a binary fashion of on or off -- transmitting light either to or away from a detector. 
If only using one detector per subspace, at any given moment a pattern may only be 
composed of 0's or 1's while Hadamard matrices are composed of 1's and -1's. To display 
the full Hadamard pattern with one detector per subspace, the data-taking operations must 
be split into positive and negative operations. $\mathbf{H}_{N^2}$ may be decomposed into 
a sum of four Kronecker products of both positive $\mathbf{H}_N$, represented by 
$\mathbf{H}^+_N$ which is composed of 0's and 1's, and negative $\mathbf{H}_N$, 
represented by $\mathbf{H}^-_N$ which is composed of -1's and 0's. \begin{equation} 
\label{eq:CorrelationMeas} \begin{split} \mathbf{H}_{N^2} = \left(\mathbf{H}_N^+\otimes 
\mathbf{H}_N^+\right) \, + \, \left(\mathbf{H}_N^-\otimes \mathbf{H}_N^-\right) \, + \, \left(\mathbf{H}_N^+\otimes 
\mathbf{H}_N^-\right) \, + \, \left(\mathbf{H}_N^-\otimes \mathbf{H}_N^+\right). \end{split} \end{equation} This 
means that every element in $\mathbf{y}$ will 
require four coincidence measurements. Even though $4M$ coincidence measurements are required when using one 
detector per subspace, the drastic sampling performance gained through CS methods is such 
that $4M\ll N^2$. Alternatively, if two detectors are used in each subspace (one detector 
to collect projections from the positive components and one detector to collect projections from the negative components), the 
detection process could be streamlined to measure each of the four correlations in 
Eq. (\ref{eq:CorrelationMeas}).

In reconstruction, fast-Hadamard transforms may be utilized by CS reconstruction 
algorithms to perform the operations $\mathbf{A}\cdot\mathbf{x}$ and $\mathbf{A}^T\cdot\mathbf{y}$. The operation 
$\mathbf{A}\cdot\mathbf{x}$ first requires that $\mathbf{x}$ be 
inverse-permuted, fast Hadamard transformed, and then have the correct $M$ elements extracted from the final result. The inverse-permutation is done by 
defining an inverse permutation vector $\mathbf{q}$ as \begin{equation}\label{eq:InvPerm} 
\mathbf{q}\left[\mathbf{p}[i]\right] = i \end{equation} for all $i$ elements in 
$\mathbf{p}$. Hence, $\mathbf{A}\cdot\mathbf{x}$ is realized with the 
following operations 
\begin{eqnarray}
\mathbf{y}' & = & 
\mathscr{H}\left[\mathbf{x}[\mathbf{q}_{SI}]\right] \nonumber \\ \mathbf{y} & = & 
\mathbf{y}'[\mathbf{r}_{SI}]. \end{eqnarray}

The operation $\mathbf{A}^T\cdot \mathbf{y}$ requires that a vector 
$\boldsymbol{\beta}$ composed of $N^2$ zeros be filled with the elements of $\mathbf{y}$ 
according to $\mathbf{r}_{SI}$, fast Hadamard transformed, and then permuted according to 
$\mathbf{q}$ as follows: 
\begin{eqnarray}\label{eq:fastAty}
\boldsymbol{\beta}[\mathbf{r}_{SI}] & = & 
\mathbf{y} \nonumber \\ \mathbf{x}[\mathbf{q}_{SI}] & = & 
\mathscr{H}\left[\boldsymbol{\beta}\right]. 
\end{eqnarray}

Again, these operations work because Hadamard matrices are symmetric. However, it should be noted that the true inverse operation is $\mathbf{H}^{-1}_{N}=\mathbf{H}^{T}_{N}/N$. When taking the fast transform operation in Eq. (\ref{eq:fastAty}), we are explicitly taking the forward fast transform and neglecting the normalization term. Because of this structure, the operations $\mathbf{A}\cdot\mathbf{x}$ and 
$\mathbf{A}^T\cdot\mathbf{y}$ can be utilized by most reconstruction algorithms to 
operate more efficiently.

The methods outlined in this article can also be applied to joint space signals that are not sparse 
in the ``pixel-basis" where $\Psi \neq \mathds{1}$. Sparse forward and inverse transform 
operations, $\Psi[...]$ and $\Psi[...]^{-1}$, need to be applied to $\mathbf{x}$ in an 
appropriate order to bring $\mathbf{x}$ and $\mathbf{y}$ back into the pixel-basis before 
fast-Hadamard transforming. Hence, the operations $\mathbf{A}\cdot\mathbf{x}$ and $\mathbf{A}^T\cdot\mathbf{y}$ become $\mathbf{A}\cdot\Psi[\mathbf{x}]^{-1}$ and $\Psi[\mathbf{A}^T\cdot\mathbf{y}]$.

To reiterate, the novelty presented in this section is in how Eqns. (\ref{eq:PicksN^2}) and (\ref{eq:PermN^2}) 
enable the use of fast Hadamard transforms for calculating $\mathbf{A}\cdot\Psi[\mathbf{x}]^{-1}$ and 
$\Psi[\mathbf{A}^T\cdot\mathbf{y}]$ as summarized below.
\begin{itemize} \item $\mathbf{y} = 
\mathbf{A}\cdot\Psi[\mathbf{x}]^{-1}$ \begin{enumerate} \item If $\Psi = \mathds{1}$, 
neglect this step. Otherwise, inverse transform $\mathbf{x}$ out of the sparse basis 
using the inverse transform to obtain $\mathbf{x}'=\Psi[\mathbf{x}]^{-1}$. \item Inverse 
permute $\mathbf{x}'$ using $\mathbf{q}_{SI}$ such that $\mathbf{x}'' = 
\mathbf{x'}[\mathbf{q}_{SI}]$. \item Fast Hadamard transform $\mathbf{x}''$ such that 
$\mathbf{y}' = \mathscr{H}[\mathbf{x}'']$. \item Extract $M$ elements from $\mathbf{y}'$ 
using $\mathbf{r}_{SI}$ to obtain $\mathbf{y} = \mathbf{y}'[\mathbf{r}_{SI}]$. 
\end{enumerate} \item $\mathbf{x} = \Psi[\mathbf{A}^T\cdot\mathbf{y}]$ \begin{enumerate} 
\item Construct a null-vector $\boldsymbol{\beta} \in \mathbb{R}^{N^2}$. \item Place the 
components of $\mathbf{y}$ into $\boldsymbol{\beta}$ using $\mathbf{r}_{SI}$ to assign 
the locations for the elements of $\mathbf{y}$ such that 
$\boldsymbol{\beta}[\mathbf{r}_{SI}] = \mathbf{y}$. \item Fast Hadamard transform 
$\boldsymbol{\beta}$ such that 
$\boldsymbol{\beta}'=\mathscr{H}\left[\boldsymbol{\beta}\right]$. \item Permute the 
elements of $\boldsymbol{\beta}'$ using $\mathbf{p}_{SI}$ to obtain 
$\mathbf{x}'=\boldsymbol{\beta}'[\mathbf{p}_{SI}]$. \item If $\Psi = \mathds{1}$, neglect 
this step and let $\mathbf{x} = \mathbf{x}'$. Otherwise, transform $\mathbf{x}'$ into the 
sparse basis to obtain $\mathbf{x} = \Psi[\mathbf{x}']$. \end{enumerate} \end{itemize}

\section{Compressive measurement in a $16.8\times 10^6$-dimensional correlated space}

\subsection{Mutual information}

To demonstrate the practicality of the previous results, we compressively measure and quickly reconstruct a $16.8\times 10^6$ dimensional joint space probability distribution. Up to this point, the reason why these joint space measurements are useful has not been explained in detail other than to inform the reader of the characterization of correlated systems. When attempting to use down-converted photons for information transfer, an important question to ask is, How much uncertainty about the position of the signal photon is removed upon knowing the position of the idler photon? This quantity is effectively answered by the Shannon mutual information between the position statistics of the signal and idler photons $I(X_S,X_I)$ \cite{dixon2012quantum}. The mutual information quantifies the classical channel capacity and is easily found by first measuring the joint space probability distribution. Given the discrete random variables' distributions for signal $X_S$ and idler $X_I$, and their allowed set of random values, $x_S$ and $x_I$ respectively, the mutual information is defined as:
\begin{equation}
I(X_S,X_I) = 
\sum\limits_{x_S \in X_S}
\sum\limits_{x_I \in X_I}
p\left(x_S,x_I\right)
\text{log}_{2}\left(\frac{p(x_S,x_I)}{p(x_S)p(x_I)}\right),
\end{equation}    
where $p(x_S,x_I)$ is the joint space probability distribution while $p(x_S)$ and $p(x_I)$ are the marginal probability distributions. In terms of our CS formalism, $p(x_S,x_I) = \mathbf{x}$. Once $\mathbf{x}$ has been procured from $\mathbf{y}$, the marginal distributions $p(x_S)$ and $p(x_I)$ are then found summing over appropriate values of $p(x_S,x_I)$. Since we may approximate the joint distribution as a double Gaussian, we may also say $2^{I(X_S,X_I)}$ is equal to the Schmidt number of the state which is a measure of the number of entangled modes \cite{laweberly2004analysis, fedorov2009gaussian}. In our case, $2^{I(X_S,X_I)}$ is equal to the number of distinct channel inputs.

\subsection{Theoretical expectations}

Before reporting our experimental results, it is useful to first estimate the theoretical maximum amount of possible mutual information as derived from first principles based on the crystal and the pump-laser specifications. That result should then be compared with the maximum possible information we could measure given our SLM resolution. A thorough calculation characterizing degenerate SPDC is done in \cite{schneeloch2015introduction} in one transverse spatial dimension. Assuming a double Gaussian bi-photon state, the mutual information in the position domain between down-converted photons ($X_S$ and $X_I$) is
\begin{equation}
\label{eq:maxmutualinfo}
I(X_S , X_I)=\text{log}_2\left(\frac{9\pi\sigma_p^2+L_z\lambda_p}{2\sigma_p\sqrt{9\pi L_z\lambda_p}}\right)
\end{equation}
where $L_z$, $\lambda_p$, and $\sigma_p$ represent the length of the nonlinear crystal, the pump-laser wavelength, and the standard deviation of the Gaussian intensity pump-laser profile, respectively.
We use a 325 nm pump laser and a 1 mm length nonlinear crystal. The maximum $1/e^2$ pump diameter is listed as 1.2 mm resulting in a $\sigma_p$ that is four times smaller, i.e. $\sigma_p = 3\times10^{-4}$ m. The experiment uses approximately degenerate down-converted light because of the paraxial nature of beam propagation and the use of narrow-band filters. Our pump laser is approximately Gaussian in two dimensions resulting in a mutual information twice as large as reported in Eq. (\ref{eq:maxmutualinfo}). We obtain a theoretical maximum mutual information between signal and idler photons of 10.9 bits. However, when moving from an infinite-dimensional Hilbert space to a finite dimensional space dictated by the resolution of the SLM and its pixel size, the resulting measurable mutual information must be less than or equal to to the continuous variable case of 10.9 bits.

\subsection{Reconstruction algorithm: maximizing the mutual information}

Given that a motivation for these reconstructions is to infer the mutual information shared between signal and idler photons, it is reasonable to design a reconstruction algorithm that attempts to maximize the mutual information via the suppression of noise through soft or hard thresholding; hard thresholding implies setting components less than a threshold to zero while soft thresholding implies first hard thresholding and then decreasing the amplitude of every remaining signal component by the threshold's amplitude. For practical applications, only the largest measurable signal components of $\mathbf{x}$ should be used to infer the mutual information because of the presence of background noise. With this in mind, we use an iterative thresholding algorithm \cite{besk2009afastiterative} that computes the mutual information at each iteration. The program exits when the mutual information no longer increases with thresholding. 

The algorithm we use is summarized as follows:
\begin{eqnarray}
\label{Eq:Algorithm}
\mathbf{x}_0 &=& \mathbf{c} \nonumber \\
\mathbf{x}_{t+1} &=& \hat{\eta}_{2}[ \mathbf{x}_t\cdot\lbrace\hat{\eta}_1\left[\mathbf{A}^T\cdot\left(\mathbf{y}-\mathbf{A}\cdot
\mathbf{x}_t\right)\right]\rbrace+\mathbf{x}_t - \text{min}\left(\mathbf{x}_t\right)]
\end{eqnarray}
where $\mathbf{c}$ is a vector composed entirely of ones times a non-zero constant and min($\mathbf{x}_t$) is a vector composed entirely of ones times the smallest element in $\mathbf{x}_t$.
Note that at each iteration we take a projection of the current result with the term $\hat{\eta}_1\left[\mathbf{A}^T\cdot\left(\mathbf{y}-\mathbf{A}\cdot
\mathbf{x}_t\right)\right]$. During the first iteration, $\mathbf{A}\cdot \mathbf{x}_0=0$ because $\mathbf{x}_0=\mathbf{c}$. Note that $\mathbf{A}\cdot \mathbf{c}=0$ only because we chose to neglect the first row of the Hadamard matrices. The first iteration results in computing $\mathbf{A}^T\cdot\mathbf{y}$ as in standard iterative style algorithms. $\hat{\eta}_1[\ldots]$ is an operator that performs soft thresholding on everything within its brackets using a biorthogonal 4.4 wavelet transform with a two-level decomposition \cite{selesnick2005dual,daubechies1990wavelet,antonini1992image}. Soft threshold in the wavelet basis, often referred to as wavelet shrinkage, is performed using the universal threshold of Donoho and Johnstone \cite{donojohnstone1994ideal,donoho1995denoising}. The filtered signal is then inverse transformed back to the pixel basis. $\hat{\eta}_2[\ldots]$ then performs a hard thresholding on everything within its brackets operating in the pixel basis and renormalizes the final result to be a probability distribution. The threshold of $\hat{\eta_2}$ gradually increases with each iteration. 
As $\mathbf{x}_t$ becomes less and less noisy through filtering and converges to the true solution, $\mathbf{y}-\mathbf{A}\cdot\mathbf{x}_t$ approaches zero. However, it never truly reaches zero and results in a small noise term. To prevent injecting random noise into each filtered iteration, we take the projection of the noisy term with the current clean solution. We then add this projection back to the current solution to prevent discarding current signal components. After hard thresholding the first iteration, $\text{min}\left({\mathbf{x}_{t>0}}\right) = \mathbf{0}$ where $\mathbf{0}$ is a null vector.

\subsection{Experimental results}

\begin{figure} \includegraphics[width=1\textwidth]{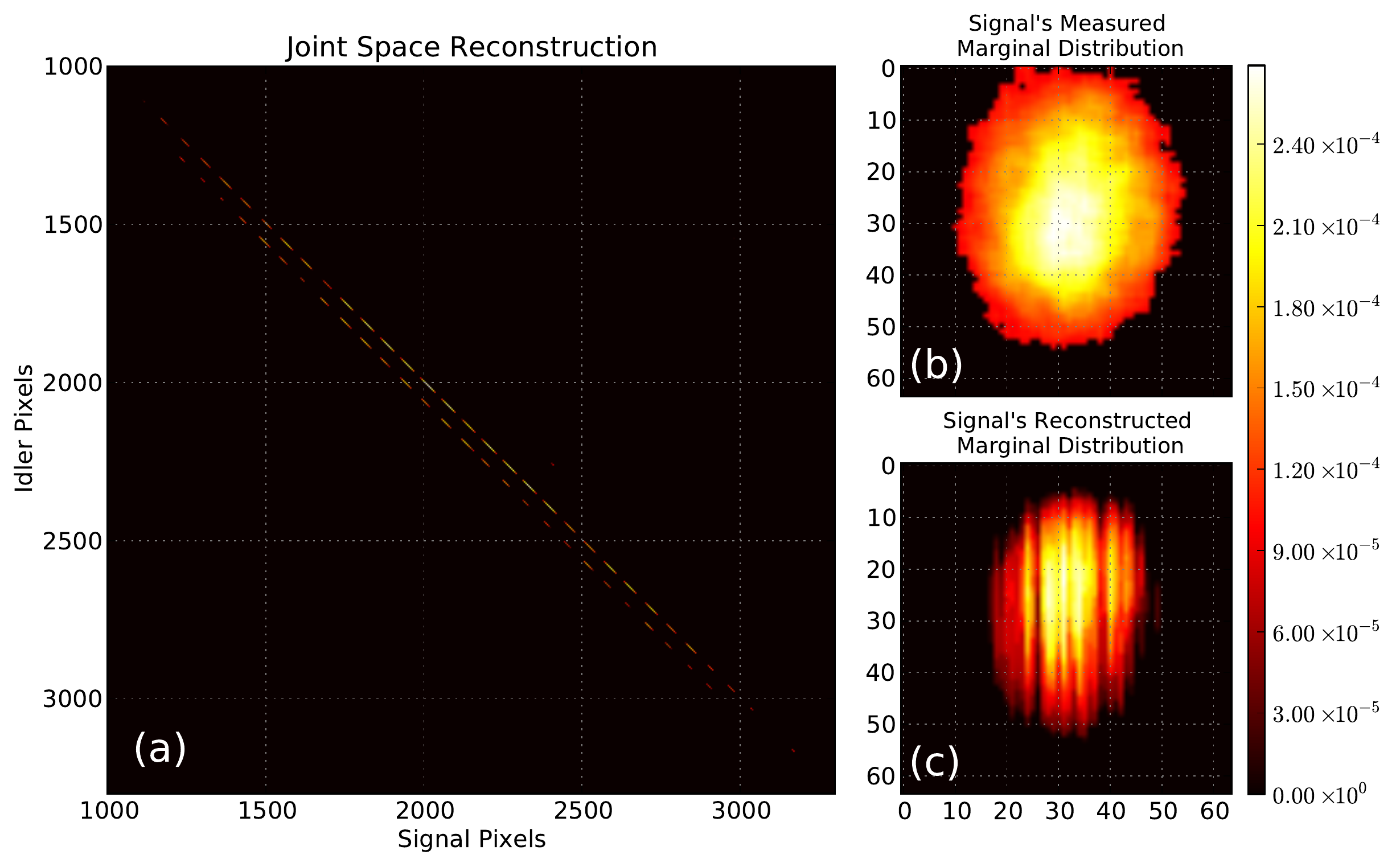} \caption{Image (a) depicts a zoomed in view that captures the largest components of the reconstructed $16.8 \times 10^6$ joint space distribution. The largest components are contained within the resulting 4-megapixel image shown. The $1/e^2$ intensity profile, as seen by the signal's SLM, is presented as the signal's measured marginal distribution in (b). The reconstructed marginal distribution was obtained through the reconstructed joint space distribution and is displayed in (c) for comparison. This data was obtained compressively with $M = 
20,000$ ($.00119\times 64^4$) samples in approximately 44 hours and was reconstructed 
in under ten minutes.} 
\label{fig:Data} \end{figure}

As an experimental demonstration, we compare how the measured mutual information from compressive measurements compares to the theoretical mutual information of 10.9 bits. Knowing that information from CS resides in the standard deviation of the signal from the different projections, this standard deviation must be greater than the shot noise. Otherwise, the signal is obscured by the noise. With binary patterns on each SLM we measured coincidence counts at a rate of $4\times 10^3$ counts per second. Since each SLM reduces the incoming flux by approximately $50\%$, the total number of coincidences $\Phi$ was approximately four times larger, or $\Phi = 1.6\times 10^4$ coincidences per second. We chose the integration time such that all four projections per $\mathbf{y}$-element required a total of 8 seconds due to power constraints. The resulting ratio of the standard deviation from the projections relative to the shot noise was 2.4. In addition, the sparsity $k$ should be of order $N = 4096$ due to tight pixel correlations resulting from position and momentum correlations, meaning $M = \mathscr{O}\left(N \text{log}(N)\right) = \mathscr{O}(10^4)$ measurements. We chose the number of measurements to be $2\times 10^4$ ($M/N^2\approx .001$) as a reasonable compromise between the total integration time and reconstruction quality. The resulting scan took just over 44 hours. To compare these values to a raster scan, the signal-to-noise ratio ($SNR$) goes as $\sqrt{\Phi \, t/N}$, assuming perfect pixel correlations and uniform illumination. Here, $t$ is the integration time per pixel. When raster scanning in an $N^2$ dimensional joint space, the total integration time goes as $N^2 t = N^3\, SNR^2/\Phi$ for shot noise limited signals. Therefore, a raster scan operating under perfect conditions, again considering perfect pixel correlations and uniform illumination, would require 50 days to achieve a $SNR = 1$ for $N = 4096$. Hence, 50 days represents the bare minimum integration time for raster scans. The reconstructed joint space probability distribution is presented in in Fig. \ref{fig:Data}. 

From the same set of data, a comparison of the recovered mutual information versus the number of samples $M$ was also conducted. The results state that the resulting mutual information is \emph{highly} dependent on the noise. As the algorithm seeks to find the optimal threshold to discern the largest number of distinguishable modes, random noise is thresholded into sparse speckle patterns when the signal is not large enough to distinguish from the noise. These speckle patterns incorrectly state that there exist tight pixel correlations. To stress the severity of this flaw, we consistently recover about 8 bits of mutual information with $M = 10$ projections. Even for large $SNR$ and large $M$, these speckle patterns may still be present along with the true signal for many reconstruction algorithms. Hence, noise will artificially increase the mutual information. 

\begin{figure} \includegraphics[width=.9\textwidth]{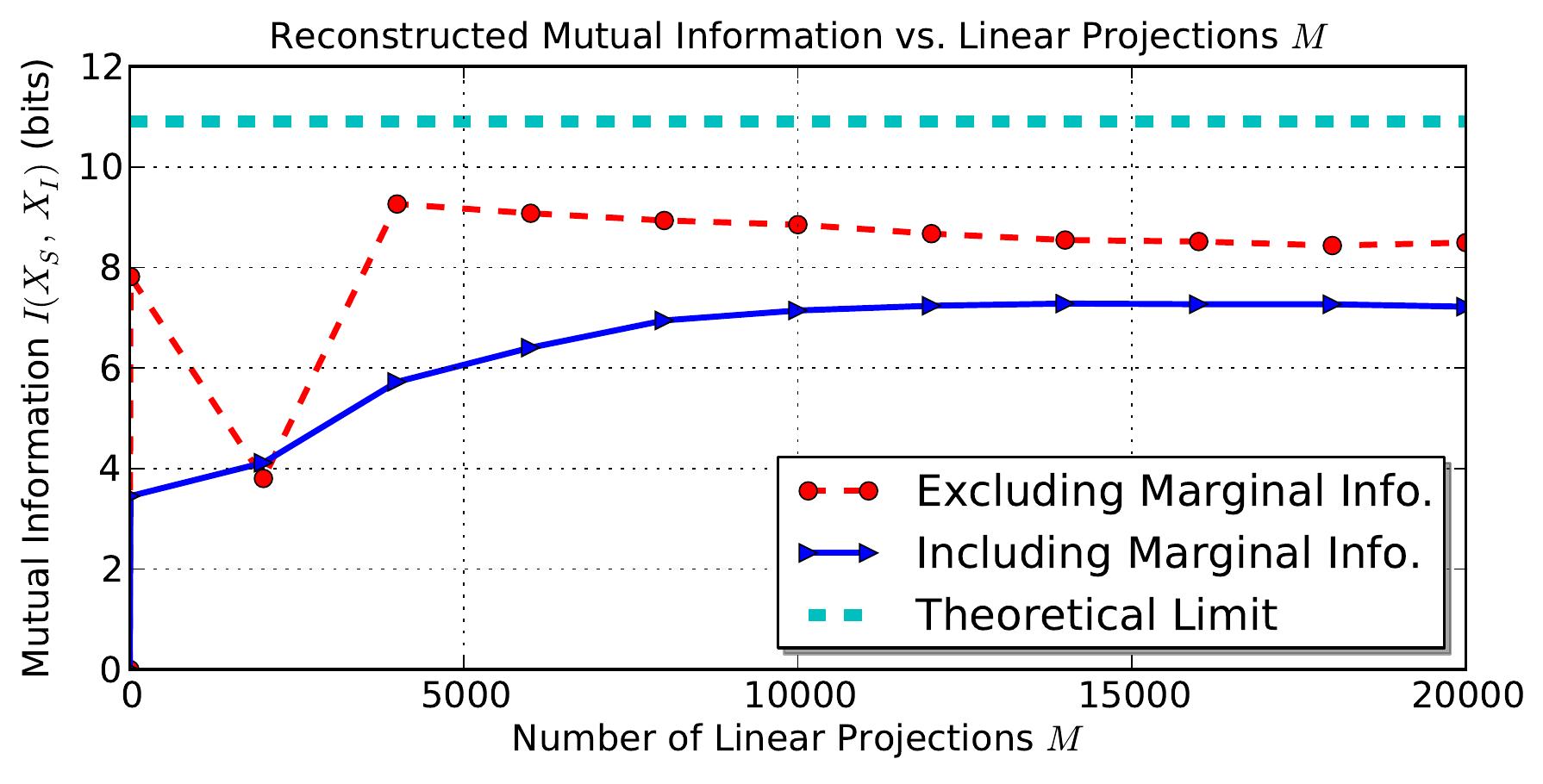}
\centering
 \caption{The mutual information, obtained through reconstructions of the joint space distribution, as a function of the number of projections $M$ is depicted above. The figure lists the theoretical maximum mutual information and the corresponding measurement results either utilizing or neglecting available information from the marginal distributions. Neglecting information found in the marginals clearly increases the inaccuracy, especially for significantly small $M$, while including this information leads to more realistic values.} 
\label{fig:MvsI} \end{figure}

Reconstruction errors that artificially increase the mutual information bring into question the validity of CS methods for these types of characterizations. However, there exists information in the signal and idler marginal probability distributions that can reduce this error. On the right hand side of Fig. \ref{fig:Data}, the marginal distribution for only the signal is shown since the idler distribution is similar in appearance. The signal's measured $64\times 64$ pixel marginal distribution was recovered from the $2\times 10^4$ projections using photon counts from only that detector. The beam is Gaussian; Fig. \ref{fig:Data} plots everything within the $1/e^2$ intensity beam diameter while thresholding the background intensity to zero. With the idler's marginal distribution taking a similar form, these marginal distributions indicate that there are regions where correlations should not exist. This information can be used to reduce the error from reconstructions by only allowing the algorithm to admit correlations where the marginal distributions say that correlations may exist. These results are summarized in Fig. \ref{fig:MvsI}. 

It is evident from Fig. \ref{fig:MvsI} that background noise significantly alters the results, fabricating a larger mutual information. When neglecting marginal information, the recovered mutual information values appear to asymptotically decrease as $M$ grows larger, supposedly approaching an accurate value. However, including information from the marginals decreases these errors, even for small $M$, by systematically reducing background noise. Note that Fig. \ref{fig:MvsI} plots the mutual information in bits. This means that for low sampling percentages when including the marginal information, just under 4 bits of mutual information, presumably from noise, is calculated. However, these reconstructions are in a regime where $M\ll k\,\text{log}(N/k)$ and the signal is dominated by noise.

Using the reconstructed joint space distribution, the marginal probability distributions can be calculated and compared to the measured marginal distributions. When not including the measured marginal information, the joint space distribution becomes riddled with a low-level noise that drastically warps the resulting reconstructed marginal distributions. However, including the marginal information results in a clean joint space distribution with reconstructed marginal distributions that are similar in appearance to the measured values. The reconstructed distributions in Fig. \ref{fig:Data} were recovered while including the marginal distribution information. Using CS and additional information in the marginal distributions, we measure 7.21 bits of mutual information, corresponding to 148 correlated modes. The reason we fall short of 10.9 bits of measured mutual information is likely due to the low resolution imposed by the beam covering a small part of the sample space. This resulted in a discretization of the marginal distributions that was not detailed enough to measure all of the available correlations. Another issue is the misalignment in the overlap of signal and idler pixels. This causes the appearance of a second correlation band as seen in Fig \ref{fig:Data} (a). Finally, a misalignment in the focus will also result in a single pixel in one arm sharing correlations with several pixels in the other arm.  

While we compressively measured the correlations existing between 4096 signal and 4096 idler SLM pixels, illumination profiles on each SLM suggest that there exist pixels where no correlations are even possible. To determine the actual joint space distribution from which we could measure possible correlations, it is reasonable to consider the $1/e^2$ diameter of each Gaussian beam profile and consider the total number of correlations that could exist between those pixels alone. The product of these areas, relative to pixel size, then defines the effective joint space dimensionality. From these enclosed areas, we calculate a joint space probability distribution of $3.23\times 10^6$ dimensions. Notice the measured and recovered marginal distributions presented in Fig. \ref{fig:Data}. The marginal distribution recovered from the reconstructed joint space distribution is smaller than the $1/e^2$ thresholded distribution displayed immediately above. These regions should be identical in the limit of infinite $SNR$ and perfect image-plane optical alignment. Again, the difference suggests that either the $SNR$ may not have been sufficient to extract all of the existing correlations or that the alignment was imprecise.

Reconstructions in \cite{howland2013efficient} were limited to a maximum resolution of $10^6$ pixels on a desktop computer with 32 GBytes of memory and required several hours to reconstruct $\mathbf{x}$ before the image quality was sufficient to exit the reconstruction program. However using a laptop with 8 GBytes of memory, we perform reconstructions of a $16.8\times 10^6$ dimensional joint space in under ten minutes with satisfactory results using fast Hadamard transforms with the algorithm presented in Eq. (\ref{Eq:Algorithm}). We ran the reconstruction algorithm 10 times per sample point to ensure that the results are consistent. The resulting error bars were four orders of magnitude smaller than the scale allows and are not shown. It should be noted that the reported mutual information values are the result of extracting the largest correlations from the data and are not the result of inferring the mutual information from a best Gaussian fit, assuming the resulting SPDC follows a commonly approximated double-Gaussian wave function \cite{laweberly2004analysis,fedorov2009gaussian}. While a Gaussian fit would probably characterize the system better and result in a higher mutual information by reporting correlations in the tails of the Gaussian, it is not indicative of the actual available channel capacity.

\section{Conclusion} This article describes in detail the methods necessary to efficiently perform 
high-dimensional compressive Kronecker imaging of a joint system. By randomizing Hadamard 
matrices and utilizing fast Hadamard transforms, CS is performed in the 16.8 
million-dimensional Kronecker space while containing a 3.23 million-dimensional bi-photon probability distribution. The 
experiment that would require over 50 days to do a raster scan under perfect conditions is instead performed in just under two days and only requires several minutes to reconstruct the data. As an example, we compressively measured 7.21 out of 10.9 bits of mutual information while also demonstrating how to improve the accuracy of these results utilizing information contained within the marginal distributions. We believe these 
methods will prove to be an invaluable tool in measuring the distribution functions of 
correlated systems as well as other correlation-based CS implementations.

\section*{Acknowledgments} We would like the thank Gregory A. Howland for the use of his 
computer code to verify our results and James Schneeloch for his contributions, theoretical discussions, and careful editing to make this a coherent article. This work was sponsored by the Air Force grant 
AFOSR Grant No. FA9550-13-1-0019.

\end{document}